\documentclass[10pt, conference, compsocconf]{IEEEtran}
\IEEEoverridecommandlockouts

\usepackage{cite}
\usepackage[pdftex]{graphicx}
\usepackage[cmex10]{amsmath}
\usepackage{array}
\usepackage[caption=false,font=footnotesize]{subfig}
\usepackage{fixltx2e}
\usepackage{url}

\usepackage{microtype}
\usepackage{verbatim}
\usepackage{color}
\usepackage{boxedminipage}
\usepackage{slashbox}
\usepackage{microtype}
\usepackage{wrapfig}

\usepackage{hyperref}

\let\labelindent\relax
\usepackage{enumitem}
\setlist{noitemsep, leftmargin=*,topsep=2pt}

\newcolumntype{C}[1]{>{\centering\arraybackslash}p{#1}}

\newcommand{\Ld}{\mathsf{Ld}}
\newcommand{\St}{\mathsf{St}}
\newcommand{\Fence}{\mathsf{Fence}}
\newcommand{\FenceLL}{\mathsf{FenceLL}}
\newcommand{\FenceLS}{\mathsf{FenceLS}}
\newcommand{\FenceSL}{\mathsf{FenceSL}}
\newcommand{\FenceSS}{\mathsf{FenceSS}}

\newcommand{\PreservePO}{<_{ppo}}
\newcommand{\ProgOrd}{<_{po}}
\newcommand{\MemOrd}{<_{mo}}
\newcommand{\ReadFrom}{\xrightarrow{r\!f}}
\newcommand{\AddrDep}{<_{adep}}
\newcommand{\DataDep}{<_{ddep}}

\newcommand{\AxiInstOrd}{InstOrder}
\newcommand{\AxiInstOrdSC}{\AxiInstOrd\textsubscript{SC}}
\newcommand{\AxiInstOrdGAM}{\AxiInstOrd\textsubscript{GAM}}

\newcommand{\AxiLdValSC}{LoadValue\textsubscript{SC}}
\newcommand{\AxiLdValGAM}{LoadValue\textsubscript{GAM}}

\newcommand{\OOOUniProc}{OOO\textsuperscript{U}}
\newcommand{\OOOMultiProc}{OOO\textsuperscript{MP}}
\newcommand{\ConstRegRAW}{RegRAW}
\newcommand{\ConstSAMemSt}{SAMemSt}

\newcommand{\ConstSAStLd}{SAStLd}
\newcommand{\ConstBrSt}{BrSt}
\newcommand{\ConstAddrSt}{AddrSt}
\newcommand{\ConstLGOrd}{LMOrd}
\newcommand{\ConstLdVal}{LdVal}
\newcommand{\ConstFence}{FenceOrd}
\newcommand{\ConstSALdLd}{SALdLd}
\newcommand{\ConstARMLdLd}{SALdLd\textsubscript{ARM}}

\newcommand{\ConstAtomLGOrd}{LMOrd\textsubscript{Atomic}}
\newcommand{\ConstAtomLdVal}{LdVal\textsubscript{Atomic}}
\newcommand{\ConstForward}{LdForward}

\newtheorem{definition}{Definition}

\begin{document}
\bstctlcite{bstctl:nodash}

\title{Constructing a Weak Memory Model\thanks{A version of this paper appears in the 45th International Symposium on Computer Architecture (ISCA), June, 2018. DOI: 10.1109/ISCA.2018.00021. \copyright 2018 IEEE.}}

\author{\begin{tabular}{ccccc}
Sizhuo Zhang\IEEEauthorrefmark{1} & Muralidaran Vijayaraghavan\IEEEauthorrefmark{1} & Andrew Wright\IEEEauthorrefmark{1} & Mehdi Alipour\IEEEauthorrefmark{2} & Arvind\IEEEauthorrefmark{1} \\
\end{tabular}\\
\begin{tabular}{ccc}
\IEEEauthorrefmark{1}MIT CSAIL & & \IEEEauthorrefmark{2}Uppsala University \\
\{szzhang, vmurali, acwright, arvind\}@csail.mit.edu & \hspace{20pt} & mehdi.alipour@it.uu.se \\
\end{tabular}}

\maketitle

\begin{abstract}
    Weak memory models are a consequence of the desire on part of architects to preserve all the uniprocessor optimizations while building a shared memory multiprocessor.
The efforts to formalize weak memory models of ARM and POWER over the last decades are mostly \emph{empirical} -- they try to capture empirically observed behaviors -- and end up providing no insight into the inherent nature of weak memory models.
This paper takes a \emph{constructive} approach to find a common base for weak memory models: we explore what a weak memory would look like if we constructed it with the explicit goal of preserving all the uniprocessor optimizations. 
We will disallow some optimizations which break a programmer's intuition in highly unexpected ways. 
The constructed model, which we call \emph{General Atomic Memory Model} (GAM), allows all four load/store reorderings.
We give the construction procedure of GAM, and provide insights which are used to define its operational and axiomatic semantics.
Though no attempt is made to match GAM to any existing weak memory model, we show by simulation that GAM has comparable performance with other models.
No deep knowledge of memory models is needed to read this paper. 
\end{abstract}


%
\IEEEpeerreviewmaketitle

\section{Introduction}\label{sec:intro}

Software programmers never asked for weak memory models.
However, they have to deal with the behaviors, which arise as a consequence of weak memory models in important commercial machines like ARM and POWER.
Many of the complications and features of high-level languages (e.g., C++11) arise because of the need to generate efficient codes for ARM and POWER, which have weak memory models~\cite{Kang:2017:PSR:3009837.3009850}.
Since the architecture community has unleashed the specter of weak memory models to the world, it should answer the following two questions:
\begin{enumerate}
    \item Do weak memory models improve PPA (performance/power/area) over strong models?
    \item Is there a common semantic base for weak memory models? This question is of practical importance because even experts cannot agree on the precise definitions of different weak models, or the differences between them.
\end{enumerate}

The first question is way harder to answer than the second one.
While ARM and POWER have weak models, Intel, which has dominated the CPU market for decades, adheres to TSO.
There are large number of papers in ISCA/MICRO/HPCA~~\cite{gharachorloo1991two,ranganathan1997using,guiady1999sc+,gniady2002speculative,ceze2007bulksc,wenisch2007mechanisms,blundell2009invisifence,singh2012end,lin2012efficient,gope2014atomic,XiangyaoTardis1,RenGpu,duan2013weefence,elver2014tso,DuanAsymmetric,Kurian,MorrisonAfek,ZhangCoup,RosTso} arguing that implementations of strong memory models may be as fast as those of weak models.
It is unlikely that we will reach consensus on this question in the short term, especially because of the entrenched interests of different companies.
Also there are no studies that we are aware of showing that one weak memory model is superior to another in terms of PPA.

This paper tries to answer the second question, i.e., find a common base for weak memory models.
Previous studies have taken an \emph{empirical} approach -- starting with an existing machine, the developers of the memory model attempt to come up with a set of axioms or rules that match the observable behavior of the machine.
However, we observe that this approach has drowned researchers in the subtly different observed behaviors on commercial machines without providing any insights into the inherent nature shared by all weak models.
For example, Sarkar et al.~\cite{sarkar2011understanding} published an operational model for POWER in 2011, and Mador-Haim et al.~\cite{mador2012axiomatic} published an axiomatic model that was proven to match the operational model in 2012.
However, in 2014, Alglave et al.~\cite{alglave2014herding} showed that the original operational model, as well as the corresponding axiomatic model, ruled out a newly observed behavior on POWER machines.
For another instance, in 2016, Flur et al.~\cite{flur2016modelling} gave an operational model for ARM, with no corresponding axiomatic model.
One year later, ARM released a revision in their ISA manual explicitly forbidding behaviors allowed by Flur's model~\cite{armv8ar}, and this resulted in another proposed ARM memory model~\cite{pulte2017simplifying}.
Clearly, formalizing weak memory models empirically is error-prone and challenging.

In this paper we take a different, a more \emph{constructive} approach to find a common base for weak memory models.
We assume that a multiprocessor is formed by connecting uniprocessors to an atomic shared memory system, and then derive the minimal constraints that all processors must obey.
We show that there are still choices left like same-address load-load orderings and dependent load orderings, each of which will result in a slightly different memory model.
Not surprisingly, ARM, Alpha and RMO differ in these choices.
Some of these choices make it difficult to specify matching operational and axiomatic definitions of the model, and give rise to confusing behaviors.
After carefully evaluating the choices, we have derived \emph{General Atomic Memory Model} (GAM).
Our hope is this insight can help architects choose a memory model before implementation and avoid spending countless hours in reverse engineering the model supported by an ISA.

We also give the formal \emph{operational} and \emph{axiomatic} definitions of GAM, which have been proven to be equivalent.
The axiomatic definition, which is a set of axioms that every legal program behavior must satisfy, can be combined with satisfiability-modulo-theory solvers to check whether a specific program behavior is allowed or disallowed~\cite{alglave2014herding,memalloy,lustig2017automated}.
The operational definition, which is an abstract machine that executes programs, is a very natural representation of actual hardware behaviors, and can be used in formal proofs based on induction for both programs~\cite{Nienhuis:2016:OSC:2983990.2983997} and hardware~\cite{ChoiKami}.

Although no attempt is made to match GAM exactly to any existing model, we show by simulation that GAM has performance comparable with other models.

In summary, this paper makes the following contributions:
\begin{enumerate}
    \item the common constraints shared by all weak memory models;
    \item GAM, a memory model based on the common constraints to avoid couner-intuitive program behaviors;
    \item the equivalent axiomatic and operational definitions of the GAM; and
    \item an evaluation showing that the performance of GAM is competitive with other weak memory models.
\end{enumerate}

\noindent\textbf{Paper organization:}
Section~\ref{sec:background} introduces the background on memory models and related works.
Section~\ref{sec:informal} shows the construction procedure of GAM.
Section~\ref{sec:formal} gives the formal definitions (i.e., axiomatic and operational definitions) of GAM.
Section~\ref{sec:sim} evaluates the performance of GAM.
Section~\ref{sec:conclude} offers the conclusion.

\section{Background and Related Works}\label{sec:background}

\subsection{Formal Definitions of Memory Models}\label{sec:background:formal}
We use SC~\cite{lamport1979make} as an example to explain the concepts of operational and axiomatic definitions.

\begin{figure}[!htb]
    \centering
    \begin{minipage}[b]{0.42\columnwidth}
        \centering
        \includegraphics[width=0.75\textwidth]{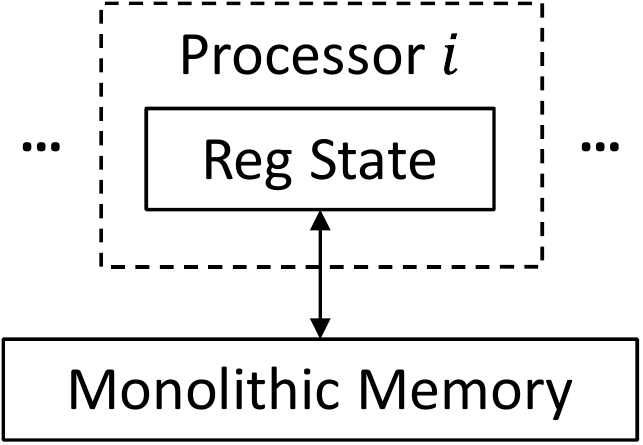}
        \vspace{-5pt}
        \caption{SC abstract machine}\label{fig:sc-op}
    \end{minipage}\hfill
    \begin{minipage}[b]{0.55\columnwidth}
        \centering
        \footnotesize
        \begin{tabular}{|l|l|}
            \hline
            \multicolumn{1}{|c|}{Proc. P1} & \multicolumn{1}{c|}{Proc. P2} \\
            \hline
            $\!\!\!I_1: \St\ [a]\ 1$          & $\!\!\!I_3: \St\ [b]\ 1$ \\
            $\!\!\!I_2: r_1 = \Ld\ [b]\!\!\!$ & $\!\!\!I_4: r_2 = \Ld\ [a]\!\!\!$ \\
            \hline
            \multicolumn{2}{|p{29ex}|}{SC allows $\langle r_1=1, r_2=1 \rangle$, $\langle r_1=0, r_2=1\rangle$ and $\langle r_1=1, r_2=0\rangle$, but forbids $\langle r_1 = 0, r_2 = 0\rangle$.}\\
            \hline
        \end{tabular}
        \caption{Litmus test Dekker}\label{fig:SB}
    \end{minipage}
\end{figure}

\noindent\textbf{Operational definition of SC:}
Figure~\ref{fig:sc-op} shows the abstract machine of SC, in which all the processors are connected directly to a monolithic memory.
The operation of this machine is simple: in one step we pick any processor to execute the next instruction on that processor \emph{atomically}.
That is, if the instruction is a reg-to-reg (i.e., ALU computation) or branch instruction, it just modifies the local register states of the processor; if it is a load, it reads from the monolithic memory instantaneously and updates the register state; and if it is a store, it updates the monolithic memory instantaneously and increments the PC.
It should be noted that no two processors can execute instructions in the same step.

As an example, consider the litmus test Dekker in Figure~\ref{fig:SB}.
If we operate the abstract machine by executing instructions in the order of $I_1\rightarrow I_2 \rightarrow I_3\rightarrow I_4$, then we get the legal SC behavior $r_1=0$ and $r_2=1$.
However, no operation of the machine can produce $r_1 = r_2 = 0$, which is forbidden by SC.

\noindent\textbf{Litmus tests:}
In the rest of the paper, we will use litmus tests like Figure~\ref{fig:SB} to show the properties of memory models or to differentiate two memory models.
A litmus test is a program snippet, and we focus on whether a specific behavior of this program is allowed by each memory model.
Since we study weak memory models, we are mostly interested in non-SC behaviors (e.g., $\langle r_1 = 0, r_2 = 0\rangle$ in Figure~\ref{fig:SB}).

\noindent\textbf{Axiomatic definition of SC:}
Before giving the axioms that program behaviors allowed by SC must satisfy, we first need to define what is a program behavior in the axiomatic setting.
For all the axiomatic definitions in this paper, a program behavior is characterized by the following three relations:
\begin{itemize}
    \item Program order $\ProgOrd$: The local ordering of instructions executed on a single processor according to program logic.
    \item Global memory order $\MemOrd$: A total order of all memory instructions from all processors, which reflects the real execution order of memory instructions.
    \item Read-from relation $\ReadFrom$: The relation that identifies the store that each load reads (i.e., store $\ReadFrom$ load).
\end{itemize}
The program behavior represented by $\langle \ProgOrd, \MemOrd, \ReadFrom \rangle$ will be allowed by a memory model if it satisfies all the axioms of the memory model.

Figure~\ref{fig:sc-axioms} shows the axioms of SC.
Axiom \AxiInstOrdSC{} says that the local order between every pair of memory instructions ($I_1$ and $I_2$) must be preserved in the global order, i.e., no rerodering in SC.
Axiom \AxiLdValSC{{} specifies the value of each load: a load can only read the youngest store among the older stores than the load in $\MemOrd$, i.e., $\max_{<mo} \{ \mathrm{set\ of\ stores} \}$.

\begin{figure}[!htb]
    \begin{boxedminipage}{\columnwidth}
        \footnotesize
        \textbf{Axiom \AxiInstOrdSC} (preserved instruction ordering):\\
        \centerline{$I_1 \ProgOrd I_2\ \Rightarrow\ I_1 \MemOrd I_2$}\\
        \textbf{Axiom \AxiLdValSC} (the value of a load):\\
        $\St\ [a]\ v \ReadFrom \Ld\ [a]\ \Rightarrow$\\
        \mbox{}\hfill$\St\ [a]\ v = \max_{<mo}\left\{\St\ [a]\ v'\ {\big|}\ \St\ [a]\ v'\MemOrd \Ld\ [a]\right\}$
    \end{boxedminipage}
    \vspace{-8pt}
    \caption{Axioms of SC}\label{fig:sc-axioms}
    \vspace{-5pt}
\end{figure}

\subsection{Atomic versus Non-atomic Memory}\label{sec:background:atomic-non-atomic-mem}
The coherent memory systems in implementations can be classified into two types: \emph{atomic} memory systems and \emph{non-atomic} memory systems, and we explain them separately.

\noindent\textbf{Atomic memory:}
For an atomic memory system, a store issued to it will be advertised to all processors simultaneously.
Such a memory system can be abstracted to a \emph{monolithic memory} which processes loads and stores instantaneously.
Implementations of atomic memory systems are well understood and used pervasively in practice.
For example, a coherent write-back cache hierarchy with a MSI/MESI protocol can be an atomic memory system~\cite{sorin2011primer,Vijayaraghavan2015}.
In such a cache hierarchy, the moment a store request is written to the L1 data array corresponds to processing the store instantaneously in the monolithic memory abstraction; and the moment a load request gets its value corresponds to the instantaneous processing of the load in the monolithic memory.

The abstraction of atomic memory can be relaxed by a little if a private store buffer is added for each processor on top of the coherent cache hierarchy.
Store buffering makes the issuing processor of a store able to see the store before any other processor, but the store still becomes visible to processors other than the issuing one at the same time.

\noindent\textbf{Non-atomic memory:}
In a non-atomic memory system, a store becomes visible to different processors at different times.
According to our knowledge, nowadays only the memory systems of POWER processors are non-atomic.
(GPUs may have non-atomic memories, but they are beyond the scope of this paper which is about CPU memory models only.)

A memory system becomes non-atomic typically because of shared store buffers or shared write-through caches.
Consider the multiprocessor in Figure~\ref{fig:shared-sb}, which contains two physical cores C1 and C2 connected via a two-level cache hierarchy.
L1 caches are private to each physical core while L2 is the shared last level cache.
Each physical core has enabled simultaneous multithreading (SMT), and appears as two logical processors to the programmer.
That is, logical processors P1 and P2 share C1 and its store buffer, while logical processors P3 and P4 share C2.
If P1 issues a store, the store will be buffered in the store buffer of C1.
In this case, P2 can read the value of the store while P3 and P4 cannot.
Besides, if P3 or P4 issues a store for the same address at this time, this new store may hit in the L1 of C2 while the store by P1 is still in the store buffer.
Thus, the new store by P3 or P4 is ordered before the store by P1 in the coherence order for the store address.
As a result, the shared store buffers together with cache hierarchy form a non-atomic memory system.

We can force each logical processor to tag its stores in the shared store buffer so that other processors do not read these stores in the store buffer.
However, if L1s are write-through caches, the memory system will be non-atomic for a similar reason, and it is impractical to tag values in the L1s.


\begin{figure}[!htb]
    \centering
    \subfloat[Shared store buffers\label{fig:shared-sb}]{\includegraphics[width=0.4\columnwidth]{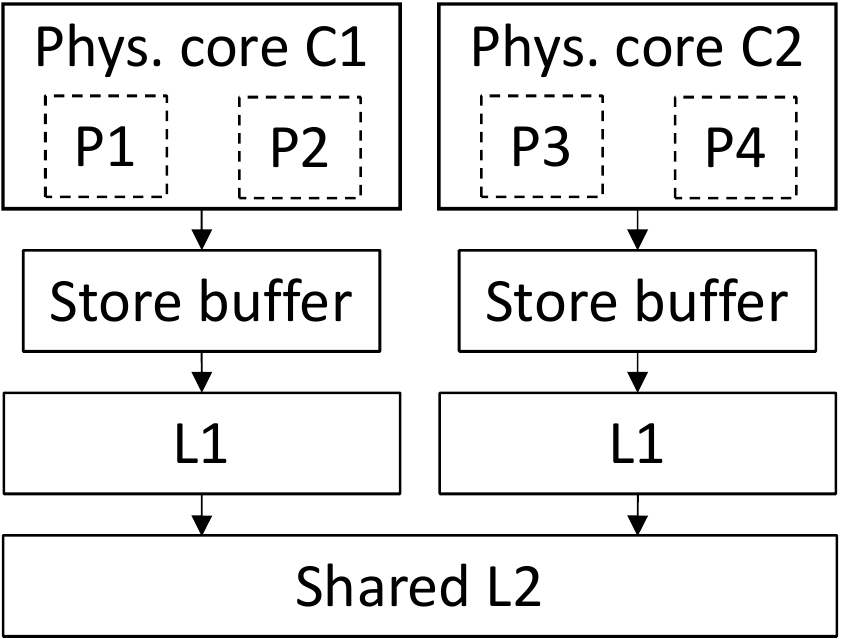}}\hspace{15pt}
    \subfloat[DASH protocol\label{fig:dash}]{\includegraphics[width=0.4\columnwidth]{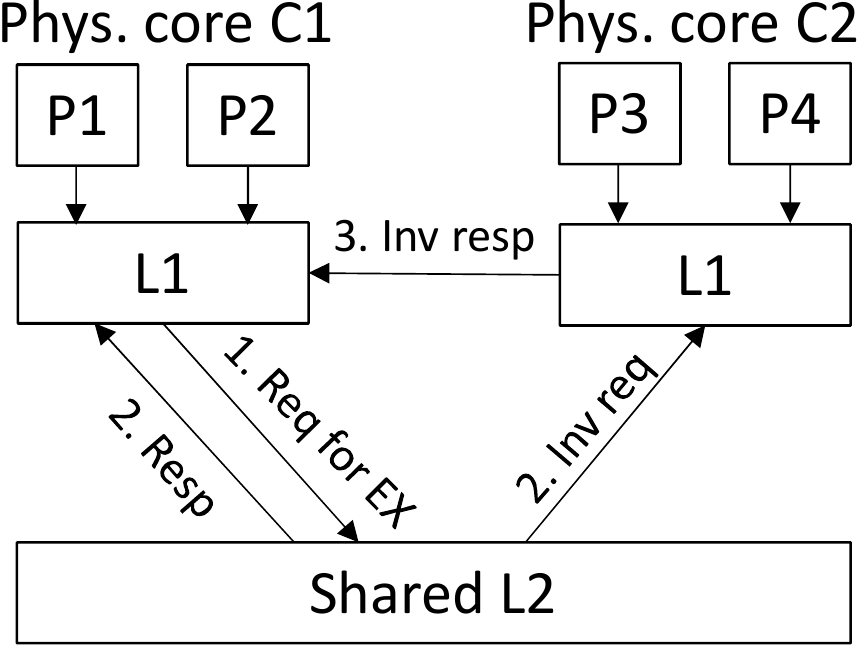}}
    \caption{Examples of non-atomic memory systems}
\end{figure}

Even if we make L1s write-back, the memory system can still fail to be an atomic memory system if it uses the DASH coherence protocol~\cite{Lenoski:1990:DCC:325164.325132} as shown in Figure~\ref{fig:dash}.
Consider the case when both L1s hold address $a$ in the shared state, and P1 is issuing a store to $a$.
In this case, the L1 of core C1 will send a request for exclusive permission to the shared L2.
When L2 sees the request, it sends the response to C1 and the invalidation request to C2 \emph{simultaneously}.
When the L1 of C1 receives the response, it can directly write the store data into the cache without waiting for the invalidation response from C2.
At this moment, P2 can read the more up-to-date store data from the L1 of C1, while P3 can only read the original memory value for $a$.
Note that in case P3 or P4 issues another store for $a$ at this moment, this new store must be ordered after the store by P1 in the coherence order of address $a$, because L2 has already acknowledged the store by P1.
This is different from non-atomic memory systems with shared store buffers or shared write-through caches.

\noindent\textbf{Atomic and non-atomic memory models:}
Because of the drastic difference in the nature of atomic and non-atomic memory systems, memory models are also classified into atomic memory model and non-atomic memory model according to the type of memory systems that the model supports in implementations.
Most memory models are atomic memory models, e.g., SC, TSO, RMO, Alpha, and ARMv8.
The only non-atomic memory model today is the POWER memory model.
In general, non-atomic memory models are much more complicated. 
In fact, ARM has recently changed its memory model from non-atomic to atomic in its version 8.
Due to the prevalence of atomic memory models, \emph{this paper discusses atomic memory models only}.

\subsection{Problems with Existing Memory Models}\label{sec:background:problem}
Here we review existing weak memory models and explain their problems.

\noindent\textbf{SC for data-race-free (DRF):}
Data-Race-Free-0 (DRF0) is an important class of software programs where races for shared variables are restricted to locks \cite{adve1990weak}.
Adve et al.~\cite{adve1990weak} have shown that the behavior of DRF0 programs is contained in SC.
DRF0 has also been extended to DRF-1~\cite{adve1993unified}, DRF-x~\cite{Marino:2010:DSE:1806596.1806636}, and DRF-rlx~\cite{SinclairRats} to cover more programming patterns.
There are also hardware schemes~\cite{ChoiDenovo,SungDenovo,SinclairSimple} that accelerate DRF programs.
While DRF is a very useful programming paradigm, a memory model for an ISA needs to specify the behaviors of all programs, including non-DRF programs.

\noindent\textbf{Release Consistency (RC):}
RC~\cite{gharachorloo1990memory} is another important software programming model.
The programmer needs to distinguish synchronizing memory accesses from ordinary ones, and label synchronizing accesses as acquire or release.
Intuitively, if a load-acquire in processor P1 reads the value of a store-release in processor P2, then memory accesses younger than the load-acquire in P1 will happen after memory accesses older than the store-release in P2.
Gharachorloo et al.~\cite{gharachorloo1990memory} have defined what is a \emph{properly-labeled} program, and shown that the behaviors of such programs are SC.

The RC definition attempts to define the behaviors for all programs in terms of the reorderings of events, and an event refers to performing a memory access with respect to a processor.
However, it is not easy to derive the value that each load should get based on the ordering of events, especially when the program is not properly labeled.
Zhang et al.~\cite{ZhangWmm} have shown that the RC definition (both RC\textsubscript{SC} and RC\textsubscript{PC}) admits some behaviors unique to non-atomic memory models, but still does not support all non-atomic memory systems in implementation (e.g., it does not support shared store buffers or shared write-through caches).

\noindent\textbf{RMO and Alpha:}
RMO~\cite{weaver1994sparc} and Alpha~\cite{alpha1998} can be viewed as variants of RC in the class of atomic memory models.
They both allow all four load/store reorderings. 
However, they have different problems regarding the ordering of dependent instructions.
RMO intends to order dependent instructions in certain cases, but its definition forbids implementations from performing speculative load execution and store forwarding simultaneously~\cite{ZhangWmm}.
Alpha is much more liberal in that it allows the reordering of dependent instructions.
However, this gives rise to  the out-of-thin-air (OOTA) problems~\cite{Boehm:2014:OGA:2618128.2618134}.

\begin{figure}[!htb]
    \centering\footnotesize
    \begin{tabular}{|l|l|}
        \hline
        \multicolumn{1}{|c|}{Proc. P1} & \multicolumn{1}{c|}{Proc. P2} \\
        \hline
        $I_1: r_1=\Ld\ [a]$  & $I_3: r_2=\Ld\ [b]$ \\
        $I_2: \St\ [b]\ r_1$ & $I_4: \St\ [a]\ r_2$ \\
        \hline
        \multicolumn{2}{|l|}{All models should forbid: $r_1=r_2=42$}\\
        \hline
    \end{tabular}
    \caption{OOTA}\label{fig:oota}
\end{figure}

Figure~\ref{fig:oota} shows an example OOTA behavior, in which value 42 is generated out of thin air.
If allowing all load/store reorderings is simply removing the \AxiInstOrdSC{} axiom from the the SC axiomatic definition, then the behavior would be legal.
To avoid such problems, Alpha introduces a complicated axiom which requires looking into all possible execution paths to determine if a younger store should not be reordered with an older load to avoid cyclic dependencies~\cite[Chapter 5.6.1.7]{alpha1998}.

To address the problems of dependencies and OOTA, the recently proposed weak memory model WMM~\cite{ZhangWmm} relaxes dependency ordering completely to avoid the complexity in specifying dependencies, but always enforces load-to-store ordering to avoid OOTA problems.
This paper, in contrast, explains where the dependency ordering constraints come from via the construction procedure of GAM; and GAM allows all four load/store reorderings. 

\noindent\textbf{ARM:}
As noted in Section~\ref{sec:intro}, ARM has been changing its memory model.
Besides, the ARM memory model also introduces complications in the ordering of loads for the same address, which we will discuss in Section~\ref{sec:informal:same-addr-ld-ld}.

\subsection{Other Related Works}

The tutorial by Adve et al. \cite{adve1996shared} has described the relations between some of the models discussed above as well as some other models \cite{goodman1991cache,dubois1986memory}.
Recently, there has been a lot of work on the programming models for emerging computing resources such as GPU~\cite{AlglaveGpu,HowerHRF,gaster2015hrf,OrrScoped,AlvarezCoherence,AlvarezRuntime}, and storage devices such as non-volatile memories~\cite{KolliLanguage,NalliPersistence,JoshiPersist,ShinPersistent}.
There are also efforts in specifying the semantics of high-level languages, e.g., C/C++~\cite{c++n4527,boehm2008foundations,batty2011mathematizing,Batty:2016:OSA:2914770.2837637,Kang:2015:FCM:2737924.2738005,Kang:2017:PSR:3009837.3009850,Nienhuis:2016:OSC:2983990.2983997} and Java~\cite{manson2005java,cenciarelli2007java, maessen2000improving}.
\emph{This paper is about CPU  memory models only.}
Model checking tools are useful in finding memory-model related bugs; \cite{alglave2014herding,LustigPipeCheck,ManerkarCCICheck,CarolineTricheck,LustigCoat,lustig2017automated} have presented tools for various aspects of memory-model testing.

\section{Intuitive Construction of GAM}\label{sec:informal}

We begin by studying a highly optimized out-of-order uniprocessor \emph{\OOOUniProc}, and show that even such an aggressive implementation still observes some ordering constraints to preserve the single-thread semantics.
When multiple \OOOUniProc{} are connected via an atomic memory system to form a multiprocessor \emph{\OOOMultiProc}, these constraints can be extended to form a base memory model that can characterize the behaviors of \OOOMultiProc{} and meet the goal of preserving uniprocessor optimizations.

However, the base model is not programmable, because there is no way to restore SC for every multithreaded program.
Therefore, we introduce fence instructions to control the exeuction order in \OOOMultiProc.
We also want to make the constructed memory model ameanable for programming, i.e., the model should not break the orderings that programmers commonly assume even when programming machines with weak memory models.
To match programmers' intuitions, we introduce more constraints to the constructed model, which means extra restrictions on implementations.
We will study the impact of these restrictions on performance in Section~\ref{sec:sim}.

\subsection{Out-of-Order Uniprocessor (\OOOUniProc)}\label{sec:informal:ooou}
Figure~\ref{fig:ooou} shows the structure of \OOOUniProc{} which is connected to a write-back cache hierarchy.
In case a memory access gets a cache miss, the processor fetches the line to L1 and then accesses it.
The memory system can process multiple requests in parallel and out of order, but will process requests for the same address in the order that they are issued to the memory system.
To simplify the description, we skip details that are unrelated to memory models.
\OOOUniProc{} fetches the next instruction speculatively, and every fetched instruction is inserted into the ROB in order.
Loads and stores will also be inserted in the same order into the load buffer (LB) and the store buffer (SB), respectively.

\begin{figure}[!htb]
    \centering
    \includegraphics[width=0.5\columnwidth]{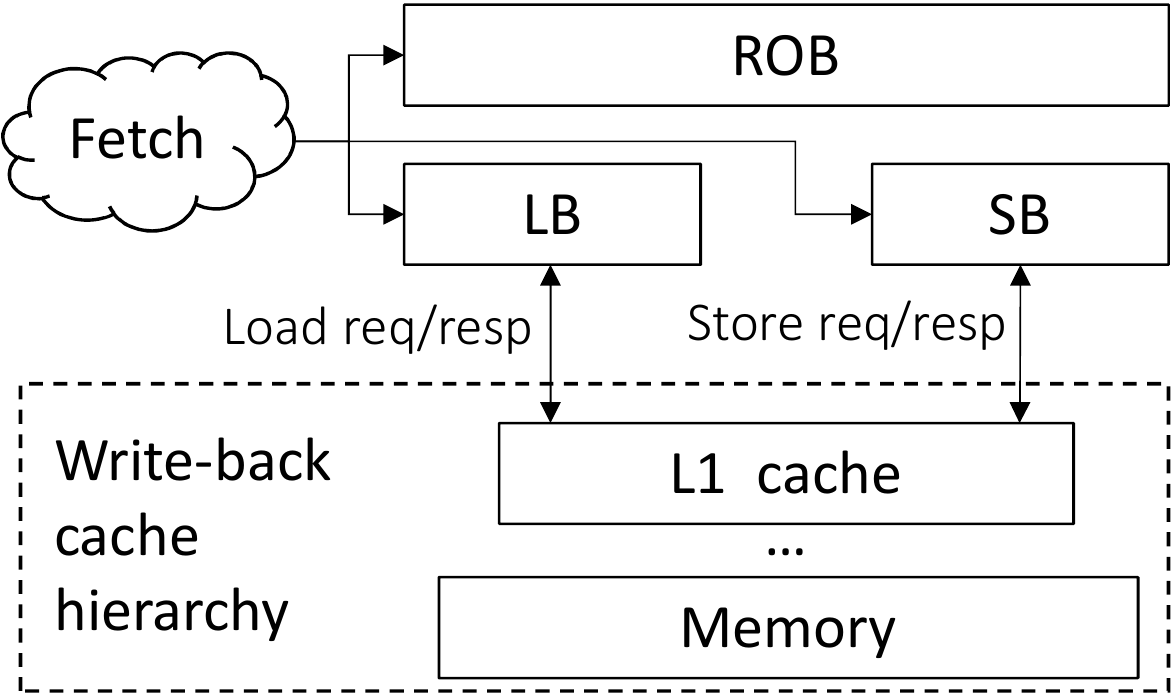}
    \caption{Structure of \OOOUniProc}\label{fig:ooou}
\end{figure}

\OOOUniProc{} executes instructions out of order and speculatively, but we assume the following two restrictions on speculations:
\begin{enumerate}
    \item A store request sent to the memory system cannot be withdrawn and its effect cannot be undone, i.e., a store cannot be sent to memory speculatively.
    \item The value of any destination register other than the next PC of an instruction is never predicted (i.e., \OOOUniProc{} does not perform any value prediction~\cite{lipasti1996exceeding,lipasti1996value,ghandour2010potential,martin2001correctly,perais2014practical,PeraisPredictor,sheikh2017load}).
\end{enumerate}
While the first restriction is easy to accept, the second one will be justified in Section~\ref{sec:dep-ld-ld}.
The restrictions on speculation imply necessary conditions when an instruction can be issued to start execution.
For example, an instruction cannot be issued until all its source operands are ready (i.e., have been computed by older instructions).
We will discuss later about other constraints on issuing an instruction (especially a store).
After being issued, a reg-to-reg or branch instruction is executed by just local computation.
The execution of a store sends a store request to the memory system.
The execution of a load first searches the SB for data forwarding from a store that has not completed the store request in the memory system.\footnote{Forwarding cannot be done after the store has been written into the L1 data array, because in the multiprocessor setting, other processors may have overwritten the value of that store.}
In case forwarding is not possible, the load will send a request to the memory system.

In spite of out-of-order execution, \OOOUniProc{} still commits instructions from ROB in order.
A store does not need to complete its store request in the memory system when being committed, while load, reg-to-reg and branch instructions should have got their values at commit time.
In the following, when we say an instruction $I_1$ is older than another instruction $I_2$, by default we mean that $I_1$ is before $I_2$ in the commit order (or equivalently, $I_1$ is inserted into ROB before $I_2$).

\noindent\textbf{Instruction reordering in the uniprocessor:}
By instruction reordering, we mean that the \emph{execution order} of two instructions is different from the \emph{commit order}.
The execution order is the order of the times when instructions \emph{finish execution}.
A reg-to-reg or branch instruction finishes execution when it computes its destination register value or resolves the next PC, respectively.
A load finishes execution when it gets forwarding from SB or reads the data from L1.
A store finishes execution when it writes the store data into the data array of the L1 cache.
An instruction that is squashed (e.g., due to mis-speculation) before being committed is not a member of the execution order.

\subsection{Constraints in \OOOUniProc}
All the constraints on the execution order in \OOOUniProc{} are listed in Figure~\ref{fig:uniproc-const}, and we will derive them one by one in the following.
These constraints can be classified into two categories.
The first set of constraints (\ConstSAMemSt{} and \ConstSAStLd) are between memory instructions for the same address, and are essential in maintaining single-thread correctness.
The second set of constraints (\ConstRegRAW, \ConstBrSt{} and \ConstAddrSt) reflects the necessary conditions that need to be met before issuing an instruction to start execution.
Although speculative execution can remove many of such conditions, some still preserve since we have assumed some restrictions on speculation.

\begin{figure}[!htb]
    \begin{boxedminipage}{\columnwidth}
        \footnotesize
        \begin{itemize}
            \item \textbf{Constraint \ConstSAMemSt} (same-address-memory-access-to-store):
            A store must be ordered after older memory instructions for the same address.
            
            \item \textbf{Constraint \ConstSAStLd} (same-address-store-to-load):
            A load must be ordered after every instruction that produce the address or data of the immediately preceding store for the same address.
            
            \item \textbf{Constraint \ConstRegRAW} (register-read-after-write):
            An instruction must be ordered after an older instruction that produce one of its source operands other than PC.
            
            \item \textbf{Constraint \ConstBrSt} (branch-to-store):
            A store must be ordered after an older branch.
            
            \item \textbf{Constraint \ConstAddrSt} (address-to-store):
            A store must be ordered after an instruction which produces the address of a memory instruction that is older than the store.
        \end{itemize}
    \end{boxedminipage}
    \caption{Constraints on execution orders in \OOOUniProc}\label{fig:uniproc-const}
\end{figure}

\noindent\textbf{Constraints for memory instructions of the same address:}
Assume $I_1$ and $I_2$ are two memory instructions for the same address $a$, and $I_1$ is older than $I_2$.
If both $I_1$ and $I_2$ are loads, then their executions do not need to be ordered.
If $I_2$ is a store, it cannot write L1 before $I_1$ finishes execution no matter whether $I_1$ is a load or a store.
Therefore we have the \emph{\ConstSAMemSt} constraint in Figure~\ref{fig:uniproc-const}.

Now consider the case that $I_1$ is a store and $I_2$ is a load.
If $I_2$ is executed by reading L1, then it cannot do so before $I_1$ has written L1.
Thus, the only way for these two instructions to get reordered is when $I_2$ gets forwarding from a store $S$ as shown in Figure~\ref{fig:store-forward}.
$S$ should be the youngest store that is older than $I_2$.
While there cannot be direct ordering constraints between $I_1$ and $I_2$ due to the forwarding, if $I_2$ eventually gets committed without being squashed, then $I_2$ cannot start execution before the address and data of $S$ have been computed by older instructions.
This gives the \emph{\ConstSAStLd} constraint in Figure~\ref{fig:uniproc-const}.

\begin{figure}[!htb]
    \centering\footnotesize
    \begin{minipage}{0.48\columnwidth}
        \centering
        \begin{tabular}{|l|}
            \hline
            \multicolumn{1}{|c|}{Proc. P1} \\
            \hline
            $I_1: \St\ [a]\ 1$ \\
            $S\ : \St\ [a]\ r_1$ \\
            $I_2: r_2 = \Ld\ [a]$ \\
            \hline
        \end{tabular}
        \caption{Store forwarding}\label{fig:store-forward}
    \end{minipage}\hfill
    \begin{minipage}{0.48\columnwidth}
        \centering
        \begin{tabular}{|l|}
            \hline
            \multicolumn{1}{|c|}{Proc. P1} \\
            \hline
            $I_1: r_1 = \Ld\ [a]$ \\
            $I_2: \St\ [r_1]\ 1$ \\
            $I_3: r_2 = \Ld\ [b]$ \\
            \hline
        \end{tabular}
        \caption{Load speculation}\label{fig:load-spec}
    \end{minipage}
\end{figure}

\noindent\textbf{Constraints for issuing to start execution:}
Since an instruction cannot be issued to execution without all its source operands being ready, we have the \emph{\ConstRegRAW} constraint in Figure~\ref{fig:uniproc-const}.
Note that we have excluded PC in this constraint.
This is because \OOOUniProc{} does branch prediction, and every fetched instruction already knows its PC and can use it for execution.

Constraint \ConstRegRAW{} has already covered the issuing requirement for reg-to-reg, branch and load instructions.
In particular, there is no more constraints regarding the issue of loads because of speculations.
For example, consider the program in Figure~\ref{fig:load-spec}.
\OOOUniProc{} can issue the load in $I_3$ before the store address of $I_2$ is computed (i.e., before $I_1$ finishes execution), even though the address of $I_2$ may turn out to be the same as $I_3$.
In case $I_1$ indeed writes value $b$ into $r_1$, \OOOUniProc{} will squash $I_3$ and re-execute it, and the execution ordering between $I_1$ and $I_3$ has been captured by constraint \ConstSAStLd.

Now we consider the constraints induced by the restriction of no speculative store issue.
A simple case is that a store cannot be issued when an older branch is not executed, i.e., constraint \emph{\ConstBrSt} in Figure~\ref{fig:uniproc-const}.
This is because the branch may be mis-predicted at fetch time and will cause an ROB squash in the future.
Another case is that a store cannot be issued when the address of an older memory instruction is not ready, i.e., constraint \emph{\ConstAddrSt} in Figure~\ref{fig:uniproc-const}.
This is because if we issue the store and later the address of the older memory instruction turns out to be same as the store address, then we may violate single-thread correctness (i.e., constraint \ConstSAMemSt).

\subsection{Extending Constraints to Multiprocessors} \label{sec:ooomp}
Consider the multiprocessor \emph{\OOOMultiProc} which connects multiple \OOOUniProc{}s to an atomic memory system which may be implemented as a coherent write-back cache hierarchy.
The constraints on local execution order in Figure~\ref{fig:uniproc-const} still apply to each \OOOUniProc{} in \OOOMultiProc, but they are not enough to describe the behaviors of the overall multiprocessor.
The only difference between a uniprocessor and a multiprocessor is about the load values.
In the uniprocessor setting, a load always gets the value of the youngest store that is older than the load.
However, in \OOOMultiProc, if a load gets its value from the atomic memory system, the value may come from a store of a different processor.

In order to understand such interaction via the atomic memory system, recall that the atomic memory system can be abstracted by a monolithic memory, and the time that a load/store request reads/writes the L1 data array in the atomic memory system corresponds to the instantaneous processing of the request in the monolithic memory (Section~\ref{sec:background:atomic-non-atomic-mem}).
Therefore, we can put all memory instructions into an \emph{atomic memory order} based on their L1 access times, which are also their execution finish times.
Hence, the atomic memory order should respect local execution order (constraint \emph{\ConstAtomLGOrd} in Figure~\ref{fig:ld-vals}), and the load that accesses the memory should read from the immediate preceding store for the same address in atomic memory order (constraint \emph{\ConstAtomLdVal} in Figure~\ref{fig:ld-vals}).

In case the load does not access the memory, it gets data forwarded from the immediate preceding store from the same processor for the same address in the commit order (constraint \emph{\ConstForward}{} in Figure~\ref{fig:ld-vals}), same as \OOOUniProc.

\begin{figure}[!htb]
    \begin{boxedminipage}{\columnwidth}
        \footnotesize
        \begin{itemize}
            \item \textbf{Constraint \ConstAtomLGOrd} (local-to-atomic-memory-order):
            The atomic memory order of two memory instructions from the same processor is the same as the execution order of these two instructions in that processor.
            
            \item \textbf{Constraint \ConstAtomLdVal} (atomic-memory-load-value):
            A load that executes by requesting the memory system should get the value of the youngest store for the same address that is ordered before the load in the atomic memory order.
            
            \item \textbf{Constraint \ConstForward} (load-forward):
            A load that executes by forwarding should get the value of the immediate preceding store from the same processor for the same address in the commit order.
        \end{itemize}
    \end{boxedminipage}
    \caption{Constraints for load values in \OOOMultiProc}~\label{fig:ld-vals}
    \vspace{-15pt}
\end{figure}

These three constraints can be restated as the two constraints \ConstLGOrd{} and \ConstLdVal{} in Figure~\ref{fig:mp-const}.
To do so, we put all memory instructions, including loads that forward from local stores, from all processors for all addresses in \OOOMultiProc{} into a \emph{global memory order} according to their execution finish times.
Thus, the global memory order should respect the atomic memory order and the execution order (constraint \emph{\ConstLGOrd}).
Note that the way a load $L$ is executed can be distinguished by the global memory order of $L$ and its immediate preceding store $S$ from the same processor for the same address in the commit order.
If $L$ is ordered before $S$ in the global memory order (i.e., $L$ finishes execution before $S$ is written to L1), then $L$ must get its value forwarded from $S$.
Otherwise, $L$ is ordered after $S$ in the global memory order, and $L$ should be executed by sending a load request to the atomic memory system.
Therefore, the constraints for load values in the two cases (\ConstAtomLdVal{} and \ConstForward) can be combined into constraint \emph{\ConstLdVal} using the following observations:
\begin{enumerate}
    \item In case of forwarding, $S$ is before $L$ in the commit order, and it is younger than (after) any store which is older than (before) $L$ in the global memory order.
    \item In case of reading the memory system, all stores that are before $L$ in the commit order are also before $L$ in the global memory order.
\end{enumerate}
Constraint \ConstLdVal{} also appears in RMO~\cite{weaver1994sparc} and Alpha~\cite{alpha1998}.

\begin{figure}[!htb]
    \begin{boxedminipage}{\columnwidth}
        \footnotesize
        \begin{itemize}
            \item \textbf{Constraint \ConstLGOrd} (local-to-global-memory-order):
            The global memory order of two memory instructions from the same processor is the same as the execution order of these two instructions in that processor.
            
            \item \textbf{Constraint \ConstLdVal} (load-value):
            A load should get the value of the youngest store for the same address in the global memory order that is ordered before the load in either the global memory order or the local commit order of the processor of the load.
        \end{itemize}
    \end{boxedminipage}
    \caption{Additional constraints in \OOOMultiProc}\label{fig:mp-const}
\end{figure}

\noindent\textbf{Atomic read-modify-write (RMW):}
There are multiple choices for the constraints that an RMW should observe.
One simple way is to say that an RMW instruction for address $a$ should obey all the constraints that apply to a load $a$ or a store $a$, and that RMW must be executed by accessing the memory system.
Due to lack of space, we do not discuss RMW in futher details in the rest of this paper.

\subsection{Constraints Required for Programming}
Up to now, the constraints in Figures~\ref{fig:uniproc-const} and \ref{fig:mp-const} are enough to describe the behaviors of loads and stores in \OOOMultiProc: constraints in Figure~\ref{fig:uniproc-const} specify which local commit order should be preserved in the local execution order, constraint \ConstLGOrd{} translates the local execution order of memory instructions to the global memory order, and finally constraint \ConstLdVal{} specifies the value of each load given the global memory order and the commit order of each processor.
However, these constraints are not enough for parallel programming especially when programmers want to restore SC.
\emph{Memory fence instructions} and \emph{enforceable dependencies} are two mechanisms to control load/store reorderings.
We will first introduce fence instructions and associated new constraints, and then discuss enforceable dependencies that have already been provided by the current constraints.
The inclusion of these new constraints results in memory model \emph{GAM0}, an initial version of GAM.

\subsubsection{Fences to Control Orderings}

Here we provide four basic fences: $\FenceLL$, $\FenceLS$, $\FenceSL$, and $\FenceSS$.
These fences order all memory instructions of a given type before the fence with all memory instructions of another given type after the fence in the execution order.
For example, $\FenceLS$ orders all loads before the fence with all stores after the fence in the execution order.
To align with our previous descriptions that each instruction has an execution finish time, we can consider that a fence also needs to be executed but acts as a NOP.
A fence restricts execution order according to the \emph{\ConstFence} (fence-ordering) constraint in Figure~\ref{fig:fence-const}.
It should be noted that a fence can only be ordered with a memory instruction, and two fences are not ordered (directly) with respect to each other.
Because of constraint \ConstLGOrd, the execution ordering enforced by fences will also apply to the global memory order.

\begin{figure}[!htb]
    \begin{boxedminipage}{\columnwidth}
        \footnotesize
        \begin{itemize}
            \item \textbf{Constraint \ConstFence} (fence-ordering):
            A $\Fence\mathsf{XY}$ must be ordered after all older memory instructions of type $\mathsf{X}$ (from the same processor) in the execution order, and ordered before all younger memory instructions of type $\mathsf{Y}$ (from the same processor) in the execution order.
        \end{itemize}
    \end{boxedminipage}
    \vspace{-6pt}
    \caption{Additional constraints for fences}\label{fig:fence-const}
\end{figure}

These fences can be combined to produce stronger fences, such as the following three which are commonly used.
\begin{itemize}
    \item Acquire fence: $\FenceLL;\ \FenceLS$.
    \item Release fence: $\FenceLS;\ \FenceSS$.
    \item Full fence: $\FenceLL;\ \FenceLS; \FenceSL; \FenceSS$.
\end{itemize}

\subsubsection{Data Dependencies to Enforce Ordering}\label{sec:dep-ld-ld}
The most commonly used enforceable dependency in programming is the data dependency.
Consider litmus test MP+addr (message passing with dependency on address) in Figure~\ref{fig:MP+addr}.
Since the address of the load in $I_5$ depends on the result of $I_4$ (i.e., $I_4$ and $I_5$ are data-dependent loads), most programmers will assume that the two loads in P2 should not be reordered, and thus the non-SC behavior $\langle r_1=a, r_2=0\rangle$ should never happen even if there is no $\FenceLL$ between the two loads in P2.
GAM0 matches this intuition of programmers because constraints \ConstRegRAW{} and \ConstLGOrd{} indeed keep $I_4$ before $I_5$ in the execution order and global memory order.

Programmers can in fact exploit the feature of data-dependent load-load ordering to replace $\FenceLL$ with artificial data dependencies.
Consider the program in Figure~\ref{fig:MP+artificial-addr}.
The intent is that P2 should execute load $b$ ($I_4$) before load $a$ ($I_6$).
To avoid inserting a fence between the two loads, one can create an artificial dependency from the result of the first load to the address of the second load.
In this way, GAM0 will still forbid the non-SC behavior.
This optimization can be useful when only $I_6$, but not any instruction following $I_6$, needs to be ordered after $I_4$, i.e., the execution of instructions following $I_6$ will not be stalled by any fence.
It should be noted that P2 should not optimize $I_5$ into $r_2=a$; otherwise there will not be any dependency from $I_4$ to $I_6$.
That is, implementations of GAM must respect \emph{syntatic} data dependency.

\begin{figure}[!htb]
    \centering
    \footnotesize
    \subfloat[MP+addr\label{fig:MP+addr}]{%
        \begin{tabular}{|l|l|}
            \hline
            \multicolumn{1}{|l|}{Proc. P1} & \multicolumn{1}{l|}{Proc. P2} \\
            \hline
            $\!\!\! I_1\!: \St\ [a]\ 1$    & $\!\!\! I_4\!: r_1 = \Ld\ [b]\!\!\!$ \\
            $\!\!\! I_2\!: \FenceSS\!\!\!$ & $\!\!\! I_5\!: r_2 = \Ld\ [r_1]\!\!\!$ \\
            $\!\!\! I_3\!: \St\ [b]\ a$    & \\
            \hline
            \multicolumn{2}{|l|}{GAM0 forbids $r_1\!=\!a, r_2\!=\!0\!\!\!$} \\
            \hline
        \end{tabular}%
    }\hfill
    \subfloat[MP+artificial-addr\label{fig:MP+artificial-addr}]{%
        \begin{tabular}{|l|l|}
            \hline
            \multicolumn{1}{|l|}{Proc. P1} & \multicolumn{1}{l|}{Proc. P2} \\
            \hline
            $\!\!\! I_1\!: \St\ [a]\ 1$     & $\!\!\! I_4\!: r_1 = \Ld\ [b]$ \\
            $\!\!\! I_2\!: \FenceSS\!\!\! $ & $\!\!\! I_5\!: r_2 = a + r_1 - r_1\!\!\!$ \\
            $\!\!\! I_3\!: \St\ [b]\ 1$     & $\!\!\! I_6\!: r_3 = \Ld\ [r_2]$ \\
            \hline
            \multicolumn{2}{|l|}{\hspace{-5pt}GAM0 forbids $r_1\!=\!1, r_2\!=\!a, r_3\!=\!0\!\!\!$} \\
            \hline
        \end{tabular}%
    }\\
    \subfloat[Dependency via memory\label{fig:dep-ld-mem}]{%
        \begin{tabular}[b]{|l|l|}
            \hline
            \multicolumn{1}{|l|}{Proc. P1} & \multicolumn{1}{l|}{Proc. P2} \\
            \hline
            $\!\!\! I_1\!: \St\ [a]\ 1$    & $\!\!\! I_4\!: r_1 = \Ld\ [b]$ \\
            $\!\!\! I_2\!: \FenceSS\!\!\!$ & $\!\!\! I_5\!: \St\ [c]\ r_1$ \\
            $\!\!\! I_3\!: \St\ [b]\ 1$    & $\!\!\! I_6\!: r_2 = \Ld\ [c]$ \\
                                           & $\!\!\! I_7\!: r_3 = a \!+\! r_2 \!-\! r_2\!\!\!$ \\
                                           & $\!\!\! I_8\!: r_4 = \Ld\ [r_3]$ \\
            \hline
            \multicolumn{2}{|p{28ex}|}{GAM0 forbids $r_1=r_2=1$, $r_3=a, r_4=0$} \\
            \hline
        \end{tabular}%
    }\hfill
    \subfloat[MP+prefetch\label{fig:MP+prefetch}]{%
        \begin{tabular}[b]{|l|l|}
            \hline
            \multicolumn{1}{|l|}{Proc. P1} & \multicolumn{1}{l|}{Proc. P2} \\
            \hline
            $\!\!\! I_1: \St\ [a]\ 1$    & $\!\!\! I_4: r_1 = \Ld\ [a]\!\!\!$\\
            $\!\!\! I_2: \FenceSS\!\!\!$ & $\!\!\! I_5: r_2 = \Ld\ [b]\!\!\!$ \\
            $\!\!\! I_3: \St\ [b]\ a$    & $\!\!\! I_6: r_3 = \Ld\ [r_2]\!\!\!$\\
            \hline
            \multicolumn{2}{|p{24ex}|}{GAM0 forbids $r_1=0$, $r_2=a, r_3=0$} \\
            \hline
        \end{tabular}
    }
    \caption{Litmus tests of data-dependency ordering}
\end{figure}

Data dependencies can not only be created by read-after-write (RAW) on registers, but also by RAW on memory locations.
GAM0 will still order two loads which are related by a chain of data dependencies via registers and memory locations.
Consider the program in Figure~\ref{fig:dep-ld-mem}.
P2 first loads from address $b$, then stores the result to address $c$, next loads from address $c$ again, and finally loads from an address $a$ which is computed using the load result on $c$.
There is a chain of data dependencies from the first load to the last load in P2, and programmers would assume that these two loads are ordered.
GAM0 indeed enforces this ordering by constraint \ConstSAStLd, which says $I_6$ should be ordered after $I_4$, i.e., the instruction that produce the data of $I_5$.

\noindent\textbf{Restrictions on implementations:}
Enforcing data-dependency ordering does not come at no cost.
As mentioned in Section~\ref{sec:informal:ooou}, the processor should not perform value prediction.
To understand why, consider again the program in Figure~\ref{fig:MP+addr}.
If P2 is allowed to perform value prediction, then it can predict the result of $I_4$ to be $a$, and issues $I_5$ to the memory system even before P1 issues any store.
This will make the non-SC behavior possible.
Martin et al.~\cite{martin2001correctly} have also noted that it is difficult to implement value prediction for weak memory models that enforce data-dependency ordering.

While value prediction is a still-evolving technique, a processor can break data-dependency ordering by just allowing a load to get data forwarding from an older executed load (i.e., load-load forwarding).
Consider the MP+prefetch litmus test in Figure~\ref{fig:MP+prefetch}.
In case load-load forwarding is allowed, P2 can first execute $I_4$ by reading 0 from memory.
Then, P1 executes all its instructions in order, and finishes writing both stores to memory.
Next P2 executes $I_5$ by reading the up-to-date value $a$ for address $b$ from memory, and finally executes $I_6$ by forwarding the stale value 0 from $I_4$.
This generates the non-SC behavior. 
To keep the data-dependency ordering, \OOOUniProc{} is only allowed to forward data from older stores as described in Section~\ref{sec:informal:ooou}.

Another technique that can break data-dependency ordering is the \emph{delayed invalidation} in the L1 cache.
That is, L1 can respond to an invalidation from the parent cache immediately without truly evicting the stale cache line.
To keep data-dependency ordering, the stale lines must be evicted if L1 is waiting for any response from the parent.
Even if the memory model does not enforce data-dependency ordering, fences have to do extra work to clear these stale lines in L1.

Enforcing data-dependency ordering is a balance between programming and processor implementation.
Nevertheless, not enforcing this ordering will result in extra fences in program patterns like pointer-chasing.
In Section~\ref{sec:sim}, we will show that forbidding load-load forwarding has negligible performance impact.
We do not evaluate the performance impact of value prediction, because it strongly depends on the effectiveness of the predictors and is beyond the scope of this paper.
Evaluation of delayed invalidation is also left to future work, because it requires appropriate multithreaded benchmarks to trigger load hits on stale lines and fence penalties to clear stale lines.

The constraints in Figures~\ref{fig:uniproc-const}, \ref{fig:mp-const} and \ref{fig:fence-const} have now formed a complete memory model, which preserves uniprocessor optimizations in implementations and has sufficient ordering mechanisms for programming.
Since this memory model targets multiprocessors with atomic memory systems, we refer to this model as \emph{General Atomic Memory Model 0} (GAM0).

\subsection{To Order or Not to Order: Same-Address Loads}\label{sec:informal:same-addr-ld-ld}

GAM0 does not have the \emph{per-location SC}~\cite{cantin2003complexity} property which many programmers expect a memory model to have.
Per-location SC requires that all accesses to a single address appear to execute in a sequential order which is consistent with the commit order of each processor.
In terms of the orderings of memory instructions for the same address, GAM0 already enforces the ordering between an older memory instruction to a younger store.
Although GAM0 allows a younger load to be reordered with an older store, the load will get the value of the store, so these two instructions can still be put into the sequential order.
The only place where GAM0 violates per-location SC is when there are two consecutive loads for the same address.
Consider the CoRR (coherent read-read) litmus test in Figure~\ref{fig:corr}.
Models with per-location SC would disallow the non-SC behavior $\langle r_1=1, r_2=0\rangle$.
However, \OOOUniProc{} can execute $I_2$ and $I_3$ out of order and there is no constraint in GAM0 to order these two loads.
Thus, the global memory order in GAM0 can be $I_3\rightarrow I_1\rightarrow I_2$, causing the non-SC behavior.
It should be noted that GAM0 is not the only memory model that violates per-location SC; RMO can also reorder two consecutive loads for the same address.

\subsubsection{Strengthen GAM0 for Per-Location SC}
To meet the programmers' requirement of per-location SC, we introduce the following \emph{\ConstSALdLd} constraint.
\begin{center}
    \begin{boxedminipage}{\columnwidth}
        \footnotesize
        \begin{itemize}
            \item \textbf{Constraint \ConstSALdLd} (same-address-load-load): The execution order of two loads for the same address (in the same processor) without any intervening store for the same address in between should match the commit order of these two loads.
        \end{itemize}
    \end{boxedminipage}
\end{center}
After introducing the above constraint to GAM0, the new memory model will forbid the non-SC behavior in Figure~\ref{fig:corr}, and we refer to the new memory model as \emph{GAM}.
Note that in constraint \ConstSALdLd, we do not order two loads with the same address in case there is a store also for the same address between them.
This is because the younger load can get forwarding from the intervening store before the older load even starts execution, and this will not violate per-location SC.
To better illustrate this point, consider the program in Figure~\ref{fig:ld-st-ld}.
$I_4$ and $I_6$ are both loads for address $b$, but there is a also a store $I_5$ for $b$ between them.
If we force $I_6$ to be after $I_4$ in the execution order and global memory order, then $I_7$ will also be ordered after $I_4$, forbidding $I_7$ from getting value 0.
However, \OOOUniProc{} can have $I_6$ bypass from $I_5$ and then execute $I_7$ by reading 0 from memory before any store in P1 has been issued.
Note that all memory accesses to $b$ can still be put into a sequential order ($I_3\rightarrow I_4\rightarrow I_5\rightarrow I_6$) which is consistent with the commit orders of P1 and P2.

To implement constraint \ConstSALdLd{} correctly, when a load resolves its address, the processor should kill younger loads for the same address which have been issued to memory or have got data forwarded from a store older than the load.
And when a load attempts to start execution, it needs to search not only older stores for the same address for forwarding but also older loads for the same address which have not started execution.
In case it finds an older load before any store, it needs to be stalled until the older load has started execution.
It should be noted that constraint \ConstSALdLd{} is a restriction on implementations purely for the purpose of matching programmers' needs.
In theory, the squashes caused by this load-load ordering constraint should affect single-thread performance.
However, in Section~\ref{sec:sim}, we will show via simulation that such squashes are very rare and the influence on performance is actually negligible.

\begin{figure}[!htb]
    \centering\footnotesize
    \subfloat[CoRR\label{fig:corr}]{%
        \begin{tabular}[b]{|l|l|}
            \hline
            \multicolumn{1}{|l|}{Proc. P1} & \multicolumn{1}{l|}{Proc. P2} \\
            \hline
            $\!\!\! I_1\!: \St\ [a]\ 1\!\!\!$ & $\!\!\! I_2\!: r_1 \!=\! \Ld\ [a]\!\!\!$ \\
                                              & $\!\!\! I_3\!: r_2 \!=\! \Ld\ [a]\!\!\!$ \\
            \hline
            \multicolumn{2}{|p{24ex}|}{Per-location SC forbids, but GAM0 and RMO allow $r_1=1, r_2=0$} \\
            \hline
        \end{tabular}%
    }\hfill
    \subfloat[Loads with an intervening store\label{fig:ld-st-ld}]{%
        \begin{tabular}[b]{|l|l|}
            \hline
            \multicolumn{1}{|l|}{Proc. P1} & \multicolumn{1}{l|}{Proc. P2} \\
            \hline
            $\!\!\! I_1\!: \St\ [a]\ 1$    & $\!\!\! I_4\!: r_1 \!=\! \Ld\ [b]$ \\
            $\!\!\! I_2\!: \FenceSS\!\!\!$ & $\!\!\! I_5\!: \St\ [b]\ 2$ \\
            $\!\!\! I_3\!: \St\ [b]\ 1$    & $\!\!\! I_6\!: r_2 \!=\! \Ld\ [b]$ \\
                                           & $\!\!\! I_7\!: r_3 \!=\! \Ld\ [a\!+\!r_2\!-\!r_2]\!\!\!$ \\
            \hline
            \multicolumn{2}{|p{30ex}|}{Both per-location SC and GAM allow $r_1=1, r_2=2, r_3=0$} \\
            \hline
        \end{tabular}%
    }\\
    \subfloat[RSW\label{fig:RSW}]{%
        \begin{tabular}{|l|l|}
            \hline
            \multicolumn{1}{|l|}{Proc. P1} & \multicolumn{1}{l|}{Proc. P2} \\
            \hline
            $\!\!\! I_1\!: \St\ [a]\ 1$      & $\!\!\! I_4\!: r_1 \!=\! \Ld\ [b]$ \\
            $\!\!\! I_2\!: \FenceSS\!\!\!\!$ & $\!\!\! I_5\!: r_2 \!=\! c \!+\! r_1 \!-\! r_1\!\!\!\!$ \\
            $\!\!\! I_3\!: \St\ [b]\ 1$      & $\!\!\! I_6\!: r_3 \!=\! \Ld\ [r_2]$ \\
                                             & $\!\!\! I_7\!: r_4 \!=\! \Ld\ [c]$ \\
                                             & $\!\!\! I_8\!: r_5 \!=\! a \!+\! r_4 \!-\! r_4\!\!\!\!$ \\
                                             & $\!\!\! I_9\!: r_6 \!=\! \Ld\ [r_5]$ \\
            \hline
            \multicolumn{2}{|p{28ex}|}{ARM allows but GAM forbids $r_1=1, r_2=c, r_3=0, r_4=0, r_5=a, r_6=0$} \\
            \hline
        \end{tabular}%
    }\hfill
    \subfloat[RNSW\label{fig:RNSW}]{%
        \begin{tabular}{|l|l|}
            \hline
            \multicolumn{1}{|l|}{Proc. P1} & \multicolumn{1}{l|}{Proc. P2} \\
            \hline
            $\!\!\! I_1: \St\ [a]\ 1$             & $\!\!\! I_4\!: r_1 \!=\! \Ld\ [b]$ \\
            $\!\!\! I_2: \FenceSS\!\!\!\!$        & $\!\!\! I_5\!: r_2 \!=\! c \!+\! r_1 \!-\! r_1\!\!\!\!$ \\
            $\!\!\! I_{10}\!\!: \St\ [c]\ 0$      & $\!\!\! I_6\!: r_3 \!=\! \Ld\ [r_2]$ \\
            $\!\!\! I_{11}\!\!: \FenceSS\!\!\!\!$ & $\!\!\! I_7\!: r_4 \!=\! \Ld\ [c]$ \\
            $\!\!\! I_3: \St\ [b]\ 1$             & $\!\!\! I_8\!: r_5 \!=\! a \!+\! r_4 \!-\! r_4\!\!\!\!$ \\
                                                  & $\!\!\! I_9\!: r_6 \!=\! \Ld\ [r_5]$ \\
            \hline
            \multicolumn{2}{|p{29ex}|}{Both ARM and GAM forbid $r_1=1, r_2=c, r_3=0, r_4=0, r_5=a, r_6=0$} \\
            \hline
        \end{tabular}%
    }
    \caption{Litmus tests for same-address loads}
\end{figure}

\subsubsection{Alternative Solution by ARM}
The ARM memory model uses a different constraint (shown below), which we refer to as \emph{\ConstARMLdLd}, to enforce the ordering of same-address loads and achieve per-location SC.
\begin{center}
    \begin{boxedminipage}{\columnwidth}
        \footnotesize
        \begin{itemize}
            \item \textbf{Constraint \ConstARMLdLd}:
            The execution order of two loads for the same address (in the same processor) that do not read from the same \emph{store} (not just same value) must match the commit order.
        \end{itemize}
    \end{boxedminipage}
\end{center}
Constraint \ConstARMLdLd{} is strictly weaker than constraint \ConstSALdLd.
To exploit the relaxation, the processor should not kill younger loads when a load resolves its address.
Instead, when a load gets its value from the memory system, the processor kills all younger loads whose values have been overwritten by other processors.
Such younger loads can be identified by keeping track of evictions from L1.
The above implementation should have less ROB squashes than the implementation of GAM with constraint \ConstSALdLd.
However, we already mentioned that the squashes in GAM are very rare, so the relaxation in constraint \ConstARMLdLd{} will not lead to extra performance.
We will confirm this point in Section~\ref{sec:sim}.

Besides little gain in performance, constraint \ConstARMLdLd{} actually gives rise to confusing program behaviors.
Consider the RSW (read-same-write) litmus test in Figure~\ref{fig:RSW} and the RNSW (read-not-same-write) litmus test in Figure~\ref{fig:RNSW}.
These two tests are very similar.
In both tests, P1 first stores to $a$ ($I_1$) and then stores to $b$ ($I_3$); P2 first loads from $b$ ($I_4$) and finally loads from $a$ ($I_9$); memory location $c$ always has value 0.
The only difference between them is that in RNSW (Figure~\ref{fig:RNSW}), P1 performs an extra store $I_{10}$ which writes the initial memory value 0 again into address $c$.
We focus on the following non-SC behavior: P2 first gets the up-to-date value 1 from $b$ ($I_4$) but finally gets the stale value 0 from $a$ ($I_9$).
Given the similarity between these two tests, one may expect that a memory model should either allow the non-SC behavior in both tests or forbid the behavior in both tests.

GAM indeed forbids this non-SC behavior in both tests, because $I_4$ and $I_6$ are data-dependent loads, $I_6$ and $I_7$ are consecutive loads for the same address $c$, and $I_7$ and $I_9$ are again data-dependent loads.
As a result, in P2, the last load must be after the first load in the global memory order in GAM, forbidding $I_9$ from getting value 0.

In contrast, ARM allows the non-SC behavior in RSW but forbids it in RNSW.
In RSW (Figure~\ref{fig:RSW}), $I_6$ and $I_7$ both reads the initial memory value and are not ordered by constraint \ConstARMLdLd{}, so the behavior is allowed by ARM.
However, in RNSW (Figure~\ref{fig:RNSW}), if $I_6$ and $I_7$ are still executed out of order to produce the non-SC behavior, then $I_7$ first reads the initial memory value and $I_6$ later reads the value of $I_{10}$.
Although the values read by $I_6$ and $I_7$ are equal, the values are supplied by different stores (initialization store and $I_{10}$), violating constraint \ConstARMLdLd.
Therefore, ARM forbids the non-SC behavior in RNSW.
We can also verify that per-location SC forbids that $I_7$ reads the initial memory value and $I_6$ reads from $I_{10}$ simultaneously, because $I_{10}$ must be ordered after the initialization of $c$ if all memory accesses for $c$ are put into a sequential order.

We believe it is confusing for constraint \ConstARMLdLd{} to allow RSW while forbidding RNSW, especially when the difference between the tests is so small.
Therefore, we resort to the much simpler \ConstSALdLd{} constraint in GAM which forbids both behaviors without losing any performance in practice.

\section{Formal Definitions of GAM}\label{sec:formal}

In this section, we give the axiomatic and operational definitions of GAM in a formal manner.
Since the axioms of GAM are similar to the constraints derived in the previous section, we give the axiomatic definition first.

\subsection{Axiomatic Definition of GAM}
\label{sec:axiomatic}

As introduced in Section~\ref{sec:background:formal}, the axiomatic definition is a set of axioms that check if a combination of program order ($\ProgOrd$), global memory order ($\MemOrd$) and read-from relation ($\ReadFrom$) is legal or not.
Program order and global memory order correspond to the commit order and the global memory order in Section~\ref{sec:informal}, respectively.
The core of the axiomatic definition of GAM is to define a \emph{preserved program order} ($\PreservePO$).
$\PreservePO$ relates two instructions in the same processor when their execution order must match the commit order.
That is, $\PreservePO$ is a summary of constraints \ConstSAMemSt, \ConstSAStLd, \ConstSALdLd, \ConstRegRAW, \ConstBrSt, \ConstAddrSt{} and \ConstFence.
After defining $\PreservePO$, we will give the two axioms of GAM, which reflect constraints \ConstLGOrd{} and \ConstLdVal{}, respectively.

Before defining $\PreservePO$, we define the RAW dependencies via registers as follows (all definitions ignore the PC register):
\begin{definition}[RS: Read Set]
    $RS(I)$ is the set of registers an instruction $I$ reads.
\end{definition}

\begin{definition}[WS: Write Set]
    $WS(I)$ is the set of registers an instruction $I$ can write.
\end{definition}

\begin{definition}[ARS: Address Read Set]
    $ARS(I)$ is the set of registers a memory instruction $I$ reads to
    compute the address of the memory operation.
\end{definition}

\begin{definition}[data dependency $\DataDep$ \label{def:data-dep}]
    $I_1 \DataDep I_2$ if $I_1 \ProgOrd I_2$ and $WS(I_1) \cap
    RS(I_2) \neq \emptyset$ and there exists a register
    $r$ in $WS(I_1) \cap RS(I_2)$ such that there is no instruction
    $I$ such that $I_1 \ProgOrd I \ProgOrd I_2$ and $r \in WS(I)$.
\end{definition}

\begin{definition}[address dependency $\AddrDep$ \label{def:addr-dep}]
    $I_1 \AddrDep I_2$ if $I_1 \ProgOrd I_2$ and $WS(I_1) \cap
    ARS(I_2) \neq \emptyset$ and there exists a register
    $r$ in $WS(I_1) \cap ARS(I_2)$ such that there is no instruction
    $I$ such that $I_1 \ProgOrd I \ProgOrd I_2$ and $r \in WS(I)$.
\end{definition}

Data dependency, i.e., $I_1 \DataDep I_2$ in Definition~\ref{def:data-dep}, means that $I_2$ will use the results of $I_1$ as its source operand.
Address-dependency, i.e., $I_1\AddrDep I_2$ in Definition~\ref{def:addr-dep}, means that $I_2$ will use the results of $I_2$ as the source operands to compute its load or store address.
Thus, data dependency includes address dependency, i.e., $I_1 \AddrDep I_2$ $\implies$ $I_1 \DataDep I_2$.

Now we define $\PreservePO$ as a summary of all the constraints for execution order:
\begin{definition}[Preserved program order $\PreservePO$\label{def:ppo}]
    Instructions $I_1 \PreservePO I_2$ if $I_1\ProgOrd I_2$ and at least one of the following is true:
    \begin{enumerate}
        \item (Constraint \ConstSAMemSt) $I_1$ is a load or store, and $I_2$ is a store for the same address.
        \item (Constraint \ConstSAStLd) $I_2$ is a load, and there exists a store $S$ to the same address such that $I_1 \DataDep S \ProgOrd I_2$, and there is no other \emph{store} for the same address between $S$ and $I_2$ in $\ProgOrd$.
        \item (Constraint \ConstSALdLd) both $I_1$ and $I_2$ are loads for the same address, and there is no store for the same address between them in $\ProgOrd$.
        \item (Constraint \ConstRegRAW) $I_1 \DataDep I_2$.
        \item (Constraint \ConstBrSt) $I_1$ is a branch and $I_2$ is a store.
        \item (Constraint \ConstAddrSt) $I_2$ is a store, and there exists a memory instruction $I$ such that $I_1 \AddrDep I \ProgOrd I_2$.
        \item (Constraint \ConstFence{} part 1) $I_1$ is a fence $\Fence\mathsf{XY}$ and $I_2$ is a memory instruction of type $\mathsf{Y}$.
        \item (Constraint \ConstFence{} part 2) $I_2$ is a fence $\Fence\mathsf{XY}$ and $I_1$ is a memory instruction of type $\mathsf{X}$.
        \item (Transitivity) \label{ppo:trans} there exists an instruction $I$ such that $I_1\PreservePO I$ and $I <_{ppo} I_2$.
    \end{enumerate}
\end{definition}
The last case in Definition~\ref{def:ppo} says that $\PreservePO$ is transitive.

With $\PreservePO$, we now give the two axioms of GAM in Figure~\ref{fig:gam-axioms}.
The \AxiLdValGAM{} axiom is just a formal way of stating constraint \ConstLdVal.
The \AxiInstOrdGAM{} axiom interprets constraint \ConstLGOrd.
That is, if two memory instructions $I_1\PreservePO I_2$, then $I_1$ should be ordered before $I_2$ in the execution order, and thus they are also ordered in the global memory order, i.e., $I_1\MemOrd I_2$.
\begin{figure}[!htb]
    \begin{boxedminipage}{\columnwidth}
        \footnotesize
        \textbf{Axiom \AxiInstOrdGAM} (preserved instruction ordering):\\
        \centerline{$I_1 \PreservePO I_2\ \Rightarrow\ I_1 \MemOrd I_2$}\\
        \textbf{Axiom \AxiLdValGAM} (the value of a load):\\
        $\St\ [a]\ v \ReadFrom \Ld\ [a]\ \Rightarrow\ \St\ [a]\ v = $ \\
        \mbox{}\hfill$\max_{<mo}\{\St\ [a]\ v'\ |\ \St\ [a]\ v'\MemOrd \Ld\ [a]\ \vee\ \St\ [a]\ v'\ProgOrd \Ld\ [a]\}$
    \end{boxedminipage}
    \vspace{-5pt}
    \caption{Axioms of GAM}\label{fig:gam-axioms}
\end{figure}

\subsection{An Operational Definition of GAM}\label{sec:formal:op}
\label{sec:operational}

The operational definition of GAM describes an abstract machine, and how to operate the machine to run a program.
Figure~\ref{fig:gam-abs-machine} shows the structure of the abstract machine.

\begin{figure}[!htb]
    \centering
    \includegraphics[width=0.4\columnwidth]{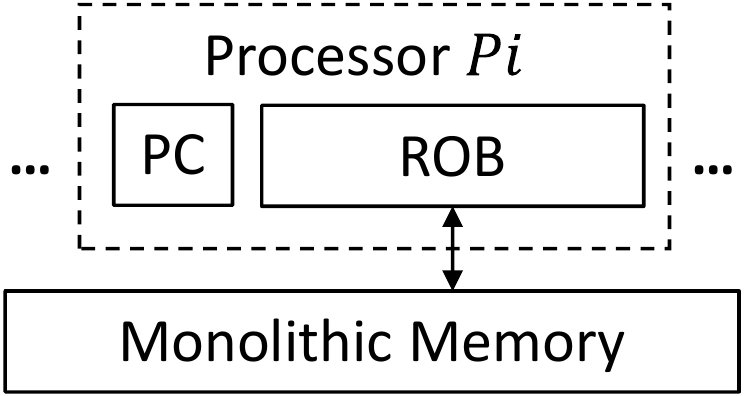}
    \vspace{-5pt}
    \caption{Abstract machine of GAM}\label{fig:gam-abs-machine}
\end{figure}

The abstract machine contains a monolithic memory (same as the one in SC) connected to each processor.
Each processor $Pi$ contains an ROB and a PC register.
The PC register contains the address of the next instruction to be fetched (speculatively) into the ROB.
The ROB has one entry per instruction; each ROB entry contains the following information for the instruction $I$ in it:
\begin{itemize}
    \item A \emph{done} bit to denote if $I$ is done or not-done (i.e., has finished execution or not).
    \item The execution result of $I$, e.g., load value or ALU result (valid only when the done bit is set).
    \item The \emph{address-available} bit, which denotes whether the memory address has been computed in case $I$ is a load or a store.
    \item The computed load or store address.
    \item The \emph{data-available} bit, which denotes if the store data has been computed in case $I$ is a store.
    \item The computed store data.
    \item The predicted branch target in case $I$ is a branch.
\end{itemize}
An instruction in ROB can search through older entries to determine if its source operands are ready and to get the source operand values.

The abstract machine runs a program in a step-by-step manner.
In each step, we can pick a processor and fire one of the rules listed in Figure~\ref{fig:gam-rules}.
That is, no two processors can be active in the same step, and the active processor in this step can fire only one rule.
Each rule consists of a \emph{guard} condition and an \emph{action}.
The rule cannot be fired unless the guard condition is satisfied.
When a processor fires a rule, it takes the action described in the rule.
The choices of the processor and the rule are arbitrary, as long as the processor state can meet the guard condition of the rule.

\begin{figure}[!htb]
    \begin{boxedminipage}{\columnwidth}
        \footnotesize
        \noindent\textbf{Rule Fetch:} Fetch a new instruction. \\
        \emph{Guard:} True. \\
        \emph{Action:} Fetch a new instruction from the address stored in the PC register.
        Add the new instruction into the tail of ROB.
        If the new instruction is a branch, predict the branch target address of the branch, update PC to be the predicted address, and record the predicted address in the ROB entry of the branch; otherwise we just increment PC.
        \vspace{2pt}\\
        \textbf{Rule Execute-Reg-to-Reg:} Execute a reg-to-reg instruction $I$. \\
        \emph{Guard:} $I$ is marked not-done and all source operands of $I$ are ready. \\
        \emph{Action:} Do the computation, record the result in the ROB entry, and mark $I$ as done.
        \vspace{2pt}\\
        \textbf{Rule Execute-Branch:} Execute a branch instruction $I$. \\
        \emph{Guard:} $I$ is marked not-done and all source operands of $I$ are ready. \\
        \emph{Action:} Compute the branch target address and mark $I$ as done.
        If the computed target address is different from the previously predicted address (which is recorded in the ROB entry), then we kill all instructions which are younger than $I$ in the ROB (excluding $I$).
        That is, we remove those instructions from the ROB, and update the PC register to the computed branch target address.
        \vspace{2pt}\\
        \textbf{Rule Execute-Fence:} Execute a $\Fence\mathsf{XY}$ instruction $I$. \\
        \emph{Guard:} $I$ is marked not-done, and for each older memory instruction $I'$ of type $\mathsf{X}$, $I'$ is done. \\
        \emph{Action:} Mark $I$ as done.
         
        \textbf{Rule Execute-load:} Execute a load instruction $I$ for address $a$. \\
        \emph{Guard:} $I$ is marked not-done, its address-available bit is set and all older $\Fence\mathsf{XL}$ instructions are done. \\
        \emph{Action:} Search the ROB from $I$ towards the oldest instruction for the first not-done memory instruction with address $a$:   
        \begin{enumerate}
            \item If a not-done load to $a$ is found then instruction $I$ cannot be executed, i.e., we do nothing.
            \item If a not-done store to $a$ is found then if the data for the store is ready, then execute $I$ by bypassing the data from the store, and mark $I$ as done; otherwise, $I$ cannot be executed (i.e., we do nothing).
            \item If nothing is found then execute $I$ by reading $m[a]$, and mark $I$ as done.
        \end{enumerate}
        \textbf{Rule Compute-Store-Data:} compute the data of a store instruction $I$. \\
        \emph{Guard:} The data-available bit is not set and the source registers for the data computation are ready. \\ 
        \emph{Action:} Compute the data of $I$ and record it in the ROB entry; set the data-available bit of the entry.
        \vspace{2pt}\\
        \textbf{Rule Execute-Store:} Execute a store $I$ for address $a$. \\
        \emph{Guard:} $I$ is marked not-done and in addition all the following conditions must be true:
        \begin{enumerate}
            \item The address-available bit of $I$ is set,
            \item The data-available bit of $I$ is set,
            \item All older branch instructions are done,
            \item \label{guard:addr->st} All older loads and stores have their address-available bits set,
            \item All older loads and stores for address $a$ are done.
            \item All older $\Fence\mathsf{XS}$ instructions are done,
        \end{enumerate}
        \emph{Action:} Update $m[a]$ and mark $I$ as done.
        \vspace{2pt}\\
        \textbf{Rule Compute-Mem-Addr:} Compute the address of a load or store instruction $I$. \\
        \emph{Guard:} The address-available bit is not set and the address operand is ready with value $a$\\
        \emph{Action:} We first set the address-available bit and record the address $a$ into the ROB entry of $I$.
        Then we search the ROB from $I$ towards the youngest instruction (excluding $I$) for the first memory instruction with address $a$. 
        If the instruction found is a done load, then we kill that load and all instructions that are younger than the load in the ROB, i.e., we remove the load and all younger instructions from ROB and set the PC register to the PC of the load.
        Otherwise no instruction needs to be killed.
    \end{boxedminipage}
    \vspace{-5pt}
    \caption{Rules to operate the GAM abstract machine}\label{fig:gam-rules}
    \vspace{-10pt}
\end{figure}

At a high level, these rules abstract the operation of processor implementations \OOOUniProc{} and \OOOMultiProc, and preserve the constraints in Section~\ref{sec:informal}.
The order of accessing monolithic memory is consistent with the global memory order in \OOOMultiProc.
Marking an instruction as done corresponds to finishing the execution of the instruction in \OOOUniProc.
Thus, the order of marking instructions as done in this abstract machine corresponds to the execution order in \OOOUniProc.
Instructions, especially loads, can be executed (i.e., marked as done) speculatively; in case this eager execution turns out to violate the constraints later on, the rules will detect the violation and squash the ROB.
Next we explain each rule.

Rule Fetch corresponds to the speculative instruction fetch in \OOOUniProc.
Rule Execute-Reg-to-Reg and Execute-Branch corresponds to finishing the execution of a reg-to-reg or branch instruction in \OOOUniProc; the guard conditions that source operands should be ready preserves constraint \ConstRegRAW.
The guard of Rule Execute-Fence preserves constraint \ConstFence.
In rule Execute-Load, the guard that checks older fences preserves constraint \ConstFence; doing nothing in case of finding a not-done load the ROB search preserves constraint \ConstSALdLd; doing nothing in case of finding a not-done store without store data preserves constraint \ConstSAStLd.
Notice that a load can be issued without waiting for all older memory instructions to resolve their addresses; this corresponds to the speculative execution in \OOOUniProc.
In the guard of rule Execute-Store, case 3 preserves constraint \ConstBrSt, case 4 preserves constraint \ConstAddrSt, case 5 preserves constraint \ConstSAMemSt, and case 6 preserves cosntraint \ConstFence.
In rule Compute-Mem-Addr, in case a store address is computed and a younger load is killed in the ROB search, constraints \ConstLdVal{} and \ConstSAStLd{} are preserved; in case a load address is computed and a younger load is killed, constraint \ConstSALdLd{} is preserved.

The proof of the equivalence of the axiomatic and operational definitions of GAM can be found in~\cite{zhang2017weak}.

\begin{figure*}[!t]
    \centering
    \includegraphics[width=\textwidth]{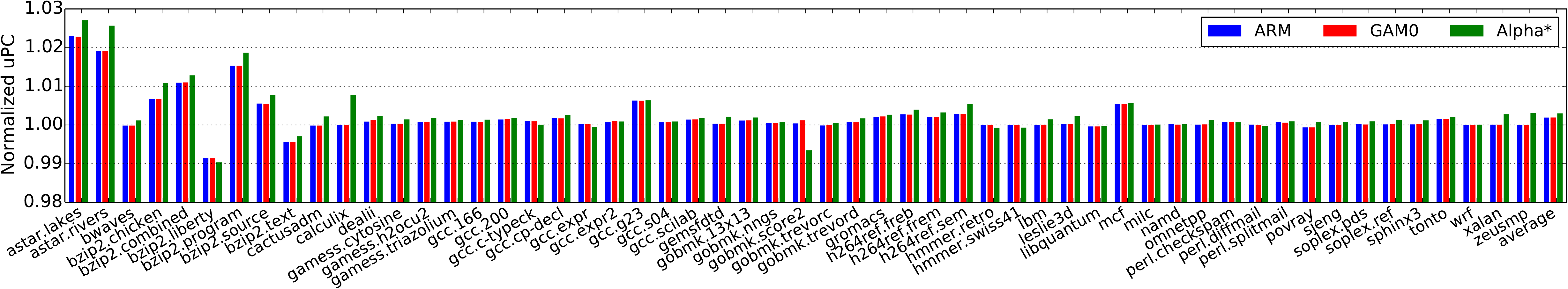}
    \vspace{-20pt}
    \caption{Normalized uPC (baseline: GAM)\label{fig:ldld-ipc}}
\end{figure*}

\section{Performance Evaluation}\label{sec:sim}

We evaluate the performance impact caused by enforcing same-address load-load ordering and disallowing load-load forwarding in GAM.

\subsection{Methodology}

As mentioned in Sections~\ref{sec:informal:same-addr-ld-ld}, the same-address load-load ordering constraint (\ConstSALdLd) places extra restrictions on uniprocessor implementations to cater to the needs of programmers.
Disallowing load-load forwarding is also mainly affecting {single-thread} performance.
Therefore, we study the performance of a \emph{single} processor of the following four memory models using the SPEC CPU2006 benchmarks:
\begin{itemize}
    \item GAM: \OOOUniProc{} with constraint \ConstSALdLd.
    \item ARM: \OOOUniProc{} with constraint \ConstARMLdLd.
    \item GAM0: \OOOUniProc{} (i.e., no constraint on same-address loads).
    \item Alpha*: \OOOUniProc{} with load-load data forwarding.
\end{itemize}
The comparison of GAM against ARM and GAM0 will show the performance impact of same-address load-load ordering constraint \ConstSALdLd, and the comparison of GAM against Alpha* will illustrate the performance implications of disallowing load-load forwarding to enforce data-dependency ordering.
As a preliminary evaluation, we do not evaluate value prediction or delayed invalidations for the reasons explained in Section~\ref{sec:dep-ld-ld}.

Besides, GAM0 can be viewed as a corrected version of RMO~\cite{weaver1994sparc} (they both allow the reordering of same-address loads).
Alpha* is similar to Alpha~\cite{alpha1998} in allowing load-load forwardings; it is more liberal than Alpha in that it does not enforce any same-address load-load ordering; but it does not account for delayed invalidations.
Thus, the comparison of GAM versus ARM, GAM0 and Alpha* will be an estimate of the performance comparison of GAM versus existing memory models including ARM, RMO and Alpha.

We modeled these four processors in GEM5~\cite{GEM5}.
The implementation details have been described in Section~\ref{sec:informal}.
For ARM, we ignore the kills when loads read values from the memory system, so the performance of ARM is an optimistic estimation.
Note that when a load is ready to issue in the ARM processor, it still searches older loads for stalls. 
Table~\ref{tab:params} shows the detailed parameters; the sizes of the important buffers (ROB, load buffer, store buffer and reservation station) are chosen to match a Haswell processor.

\begin{table}[!htb]
    \centering
    \footnotesize
    \caption{Processor parameters}\label{tab:params}
    \begin{tabular}{|p{10ex}|p{50ex}|}
        \hline
        \multicolumn{2}{|c|}{Single core @2.5GHz with x86 ISA (modified O3 CPU model)} \\ \hline
        Width & 4-way fetch/decode/rename/commit, 6-way issue to execution, 6-way write-back to register file \\ \hline
        Function units & 4 Int ALUs, 1 Int multiply, 1 Int Divide, 2 FP ALUs, 1 FP multiply, 1 FP divide and sqrt, 2 load/store units \\ \hline
        Buffers & 192-entry ROB, 60-entry reservation station, 72-entry load buffer, 42-entry store buffer (holding both speculative and committed stores) \\ \hline
        \multicolumn{2}{|c|}{Classic memory system with 64B cache lines} \\ \hline
        L1 inst & 32KB, 8-way, 4-cycle hit latency, 4 MSHRs \\ \hline
        L1 data & 32KB, 8-way, 4-cycle hit latency, 8 MSHRs \\ \hline
        Unified L2 & 256KB, 8-way, 12-cycle hit latency, 20 MSHRs \\ \hline
        L3 & 1MB, 16-way, 35-cycle hit latency, 30 MSHRs \\ \hline
        Memory & 80ns (200-cycle) latency and 12.8GB/s bandwidth \\ \hline
    \end{tabular}
    
\end{table}

We run all reference inputs of all SPEC CPU benchmarks (55 inputs in total) in full-system mode.
For each input, we simulate from 10 uniformly distributed checkpoints.
For each checkpoint, we first warm up the memory system for 25M instructions, then warm up the processor pipeline for 200K instructions, and finally simulate 100M instructions in detail.
We summarize the statistics of the 10 checkpoints to produce the final performance numbers for this benchmark input.

Since GEM5 cracks an instruction into micro-ops (uOPs), we will use uOP counts instead of instruction counts when reporting performance numbers.

\subsection{Results and Analysis}

Figure~\ref{fig:ldld-ipc} shows the uOPs per cycle (uPC) of ARM, GAM0 and Alpha* for each benchmark input.
The numbers are normalized against the uPC of GAM.
The last column in the figure is the average across all benchmark inputs.
As we can see, the performance improvements of ARM, GAM0 and Alpha* over GAM are all negligible (less than 0.3\% on average) and never exceed 3\%.
This shows that the performance penalty for GAM to enforce the same-address load-load ordering and data-dependency ordering is very small.
Next we analyze the influence of these two orderings in more details.

\noindent\textbf{Same-address load-load ordering:}
Constraint \ConstSALdLd{} in GAM puts the following two restrictions on implementations:
\begin{enumerate}
    \item \textbf{Kills}: when a load $L$ computes its address, the processor kills any younger load which has finished execution but has not got its value from a store younger than $L$.
    \item \textbf{Stalls}: when a load $L$ is ready to issue to start execution, if there is an older unissued load for the same address and $L$ cannot get forwarding from any store younger than the unissued load, then $L$ will be stalled.
\end{enumerate}
In contrast, ARM will not have any kills, but it is still subject to the stalls; GAM0 is not affected by the kills or the stalls.
Table~\ref{tab:ldld-kills-stalls} shows the number of kills or stalls (caused by same-address load-load ordering) per thousand uOPs in GAM and ARM.
Due to lack of space, we just show the maximum and average numbers across all benchmarks.
As we can see, both kills and stalls caused by same-address load-load orderings are very rare, so GAM can still have competitive performance.

\begin{table}[!htb]
    \centering
    \footnotesize
    \caption{Kills and stalls caused by same-address load-load ordering in GAM and ARM}\label{tab:ldld-kills-stalls}
    \begin{tabular}{|c|C{15ex}|C{15ex}|}
        \hline
        & \multicolumn{2}{c|}{Number of events per 1K uOPs} \\ \cline{2-3}
                      & Average & Max \\ \hline\hline
        Kills in GAM  & 0.2     & 3.24 \\ \hline
        Stalls in GAM & 0.19    & 2.15 \\ \hline
        Stalls in ARM & 0.19    & 2.15 \\ \hline
    \end{tabular}
    
\end{table}

\noindent\textbf{Load-Load forwarding:}
In case data-dependency ordering is not enforced, the processor (i.e., Alpha*) can forward data from an older executed load to a younger unexecuted load.
However, this forwarding is beneficial only in case that the younger load would get a cache miss if it were issued to the memory system.
Table~\ref{tab:ldld-forwards} summarizes the effectiveness of load-load forwardings in Alpha* (only the average and maximum numbers across all benchmarks are shown here).
The first row shows the number of load-to-load forwardings per thousand uOPs in Alpha*.
The second row shows the reduction of Alpha* over GAM in the number of L1 load misses per thousand uOPs.
As we can see, load-load forwardings can happen quite frequently.
However, the number of L1 load misses is not reduced, i.e., the load that gets the forwarding from the older load can also read the data from the L1 cache.
This explains why load-load forwardings do not translate to performance improvement over GAM.

\begin{table}[!htb]
    \centering
    \footnotesize
    \caption{Effects of load-load forwardings in Alpha*}\label{tab:ldld-forwards}
    \begin{tabular}{|c|C{11.5ex}|C{11.5ex}|}
        \hline
                                        & \multicolumn{2}{c|}{Number of events per 1K uOPs} \\ \cline{2-3}
                                        & Average & Max \\ \hline\hline
        Load-load forwardings           & 22      & 104 \\ \hline
        Reduced L1 load misses over GAM & 0.01    & 0.73 \\ \hline
    \end{tabular}
\end{table}

\section{Conclusion}\label{sec:conclude}

This paper constructed a weak memory model, GAM, which preserves all uniprocessor optimizations except those breaking programmers' intuitions.
The construction of GAM starts from the constraints on execution orders in uniprocessors, then extends the constraints to a multiprocessor setting, and finally introduces additional constraints necessary for parallel programming.
This construction procedure makes GAM a memory model that preserves most uniprocessor optimizations.
It also explains why each ordering constraint is introduced, and which uniprocessor optimizations are sacrificed for programming purposes. 
Our evaluation shows that the performance of GAM is comparable to other weak memory models.
In particular, the number of kills and stalls caused by enforcing same-address load-load ordering are negligible.


\section*{Acknowledgment}

We thank all the anonymous reviewers and especially our shepherd Thomas Wenisch for their helpful feedbacks on improving this paper.
We have also benefited from the help from Martin Rinard, Thomas Bourgeat, Daniel Lustig, and Trevor Carlson.
This work was done as part of the Proteus project under the DARPA BRASS Program (grant number 6933218).

\bibliographystyle{IEEEtran}
\def\IEEEbibitemsep{0pt}
\bibliography{short}

\end{document}